\documentclass[journal=langd5,manuscript=letter,layout=onecolumn]{achemso}
\setkeys{acs}{articletitle=true}

\usepackage{natbib}
\usepackage{graphicx}
\usepackage{multirow}
\usepackage{placeins}
\usepackage{float}
\usepackage{amsmath}
\usepackage{color}
\usepackage{amssymb}
\usepackage{hhline}
\usepackage[normalem]{ulem}
\usepackage{natbib}

\usepackage[version=3]{mhchem} % Formula subscripts using \ce{}
\usepackage[T1]{fontenc}

\newcommand*\diff{\mathop{}\!\mathrm{d}}

\newcommand\abs[1]{\left|#1\right|}

\author{Arash~Azari} \email {arash.azari@gmail.com}
\author{J\'{e}r\^{o}me~J. Crassous}\email {jerome.crassous@fkem1.lu.se}
\author{Adriana~M. Mihut}
\author{Erik~Bialik}
\author{Peter~Schurtenberger}
\author{Joakim~Stenhammar}
\author{Per~Linse}

\affiliation{Physical Chemistry, Department of Chemistry, Lund University, SE-22100 Lund, Sweden}

\title{Directed Self-Assembly of Polarizable Ellipsoids in an External Electric Field}

\abbreviations{DSA, BCT, FCT}

\begin{document}

%\begin{tocentry}
%\begin{center}
%\includegraphics[width=60mm]{TOC3.pdf}
%\end{center}
%\end{tocentry}

\begin{abstract}
The interplay between shape anisotropy and directed long-range interactions enables the self-assembly of complex colloidal structures. As a recent highlight, ellipsoidal particles polarized in an external electric field were observed to associate into well-defined tubular structures. In this study, we investigate systematically such directed self-assembly using Monte Carlo simulations of a two-point-charge model of polarizable prolate ellipsoids. In spite of its simplicity and computational efficiency, we demonstrate that the model is capable of capturing the complex structures observed in experiments on ellipsoidal colloids at low volume fractions. We show that, at sufficiently high electric field strength, the anisotropy in shape and electrostatic interactions causes a transition from 3-dimensional crystal structures observed at low aspect ratios to 2-dimensional sheets and tubes at higher aspect ratios. Our work thus illustrates the rich self-assembly behavior accessible when exploiting the interplay between competing long- and short-range anisotropic interactions in colloidal systems.
\end{abstract}

\section{Introduction}
%{\bf {\blue Please use the second \verb {\documentclass}  for single column style.Please add/change words to/in the abbreviations and keywords lists. }}

Mesoscopic self-assembly is a key principle underlying all biological systems, taking place in for example membrane formation, DNA packing and the assembly of cytoskeletal filaments~\cite{whitesides2}. Partially inspired by these principles, the self-assembly of synthetic colloidal particles has become a very active research field over the last decades, both because of their suitability as simple model systems and due to their potential applications as building blocks for functional materials~\cite{whitesides2,grzelczak,vogel,karg,dong,glotzer,sacanna,kim,tavacoli}.

The colloidal self-assembly process can be tuned both through the colloidal design, \emph{i.e.}, the size, shape and material properties of the building blocks, and by controlling the magnitude of the interparticle interactions. In addition to these control parameters, the application of external stimuli, through for example external fields, fluid flows, or patterned surfaces, enables further manipulation and control of the resulting structures through the so-called \emph{directed self-assembly} (DSA)~\cite{grzelczak}. One such example is the self-assembly of polarizable colloids using an external electric~\cite{dassanayake,blair,yethiraj,rotunno,hynninen,Agarwal2009,Dobnikar2013,Sofi,Jerome,barros,troppenz,Per2,Jerome2,Mohanty2015PRX}  or magnetic~\cite{Smallenburg2012,Wang2013,Swan2014,Reinink2014,Donaldson2017Nanoscale,Donaldson2017ACSNano,Martchenko2016,malik} field. The application of a uniform external field causes a net polarization of the colloids, which for a single spherical particle is exactly described by an ideal dipole placed at the center of the particle~\cite{Bottcher}. The interaction between the induced dipole moments will thus lead to the formation of higher order structures such as dipolar strings and, for higher densities and field strengths, networks and crystals~\cite{dassanayake,yethiraj,hynninen,Agarwal2009,Dobnikar2013,Sofi,Mohanty2015PRX}. For non-spherical colloids, such as spherocylinders and ellipsoids, the phase diagram becomes even richer due to the interplay between electro- or magnetostatic interactions, which now need to be described by including higher-order terms in the multipole expansion, and orientation-dependent excluded volume interactions \cite{blair,rotunno,Jerome,troppenz,Jerome2}. In particular, it was recently shown that prolate ellipsoids could reversibly assemble into well-defined microtubules under the application of an AC electric field.\cite{Jerome}

Due to the complexity of electrostatic interactions between anisotropic polarized bodies, the phase behavior of such colloids remains challenging to explore. A few studies have so far computationally investigated DSA of non-spherical particles such as spherocylinders~\cite{rotunno,troppenz}, ellipsoids~\cite{blair,Jerome} and ``superballs''~\cite{Per2} using external electric or magnetic fields. While ellipsoidal particles are geometrically rather close to spherocylinders, the tubular phase observed for ellipsoids does not appear in the latter case.~\cite{liu,troppenz} This indicates an intricate interplay between shape anisotropy and directional electrostatic interactions. In this study, using Monte Carlo simulations, we report a systematic numerical investigation of DSA of polarized ellipsoids in an external electric field at relatively low volume fractions $\phi \leq 0.1$, corresponding to an experimentally relevant parameter range. We discuss the use of different particle models and compare their two-body energy landscapes, demonstrating the crucial role played by electric moments higher than the dipole when dealing with non-spherical colloids, as using a purely dipolar potential gives erroneous results for anything but small aspect ratios. We then present the simulated state diagram as a function of aspect ratio and field strength and compare it to the experimental observations made in Ref.~\cite{Jerome}~. We show that all the experimentally observed structures (1-dimensional strings, 2-dimensional sheets and tubes, and 3-dimensional crystals and aggregates) are reproduced using a simplified model of polarized ellipsoids, consisting of only two point charges of properly adjusted magnitude and separation. Our results highlight the rich self-assembly behavior accessible when exploiting the combination of orientation-dependent excluded volume and long-range electrostatic interactions.\\

\section{Model description and validation}\label{sec:model}

\begin{figure}[t]
\includegraphics[width=0.9\columnwidth]{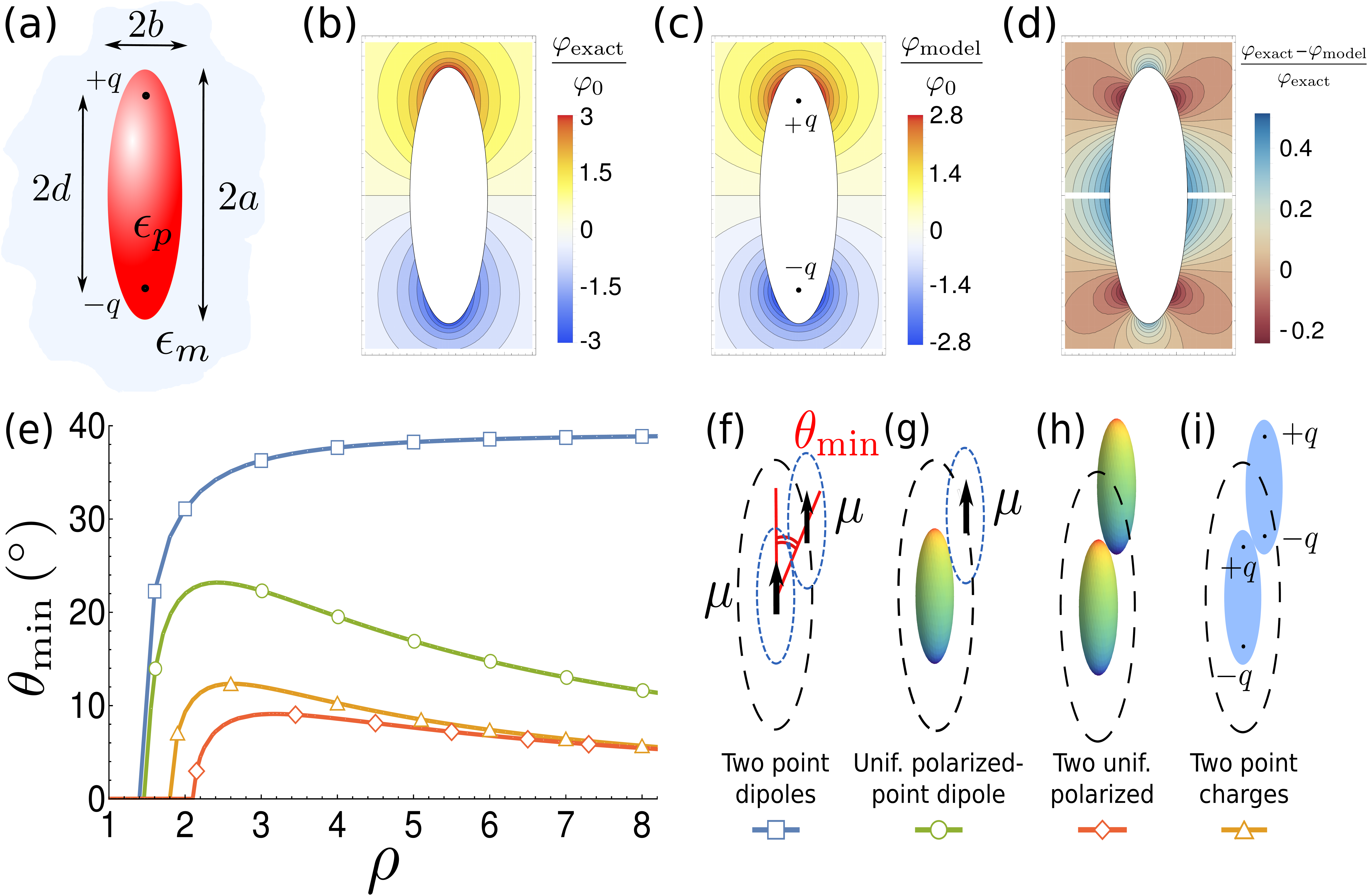}
\caption{(a) Schematic representation of the two-charge model used in the simulations. (b, c) Electrostatic potential $\varphi$, normalized by the external potential $\varphi_0$, of (b) a uniformly polarized ellipsoid (Eq. \eqref{pot_out}) and (c) the corresponding two-charge model. (d) Relative difference $(\varphi_{\rm exact}-\varphi_{\rm model})/\varphi_{\rm exact}$ between the potential of a uniformly polarized ellipsoid and that of the two-charge model. All ellipsoids have the same aspect ratio, $\rho=3.3$. (e) Angle $\theta_{\mathrm{min}}$ between two adjacent parallel ellipsoids at their minimum-energy configuration as a function of their aspect ratio $\rho$ using the four models described in panels (f-i) and in the text. Note that the symbols in (e) do not represent the full set of data points, but are for labelling purposes only.} \label{figure1}
\end{figure}

We consider a prolate ellipsoidal particle with long axis $a$ and short axes $b=c$ made of a dielectric material with dielectric permittivity $\epsilon_p$ immersed in a medium of dielectric permittivity $\epsilon_m$; see Fig.~\ref{figure1}\textbf{a}. The system is subjected to an external electric field $\mathbf{E}_0$ of magnitude $E_0$. Due to the difference in dielectric permittivities between the particle and the medium, the particle becomes polarized and aligns its long axis with the external field. In ellipsoidal coordinates $(\xi, \eta, \zeta)$, the electrostatic potential outside the ellipsoid resulting from the polarization charge density can be expressed as~\cite{stratton}

\begin{equation}
\varphi_{\rm out}= \varphi_{0} \frac{ {\displaystyle \int^{\xi} \frac{\diff s}{ \left( s+a^{2}    \right) f(s) }  } }{ {\displaystyle \frac{2 \epsilon_{m}}{abc \left( \epsilon_{p}-\epsilon_m \right) }} + {\displaystyle \int_{0}^{\infty} \frac{\diff s}{ \left( s+a^{2}    \right) f(s)  }   }   },    \label{pot_out}
\end{equation}

\noindent where $\varphi_{0}$ is the unperturbed external potential and $f(s) = \sqrt{\left(s + a^{2} \right) \left(s+ b^{2} \right) \left(s+ c^{2} \right)}$. The electrostatic potential map around a polarized ellipsoid with aspect ratio $\rho = a/b = 3.3$ is shown in Fig.~\ref{figure1}\textbf{b}. Directly using the potential of Eq.~\eqref{pot_out} for simulating a collection of particles quickly becomes computationally demanding, especially if many-body polarization between particles is taken into account. Thus, we now adopt a model where the potential in Eq.~\eqref{pot_out} is approximated by the potential stemming from two opposite point charges $\pm q$ separated by a distance $2d$ (Fig.~\ref{figure1}\textbf{a}), as has previously been adopted for the modeling of polarized spherocylinders~\cite{rotunno,troppenz} and ellipsoids~\cite{Jerome}. In an AC electric field, the polarization of the particles is predominantly determined by the conduction of counterions in or at the surface of the microgel layer and thus follows the Maxwell-Wagner-O'Konski mechanism~\cite{morgan,Jerome}. At the experimental AC frequency of Ref.~\cite{Jerome}~ (160 kHz), the interactions are treated as unscreened electrostatic interactions~\cite{Jerome} as the very dilute background electrolyte does not have time to respond at this frequency.

As shown in Fig.~\ref{figure1}\textbf{c,d} and in Supporting Information (SI), by properly adjusting $d$ and $\abs q$, a fairly accurate description of the exact potential for moderate aspect ratios (see SI) can be achieved. The charge separation $d$ was adjusted to match the ratio between the two lowest non-vanishing multipole moments (the dipole and the octupole) of the exact solution, yielding ${\displaystyle d^{2}= 0.6 b^{2} \left( \rho^{2} -1 \right)}$~\cite{schult,Jerome}. Note that, apart from using an approximation for the potential of Eq.~\eqref{pot_out}, we furthermore assume that (\emph{i}) the particles are fully aligned in the electric field, $i.e.$, the coupling with the field is not explicitly included in the simulations, and (\emph{ii}) that the many-body polarization between particles is neglected and the charge distribution of the particles remains fixed.

In order to validate our model and illustrate the importance of shape anisotropy at the two-particle level, in Fig.~\ref{figure1}\textbf{e} we evaluate the contact angle $\theta_{\mathrm{min}}$ between two adjacent ellipsoids with fixed orientation at their minimum energy configuration as a function of their aspect ratio $\rho$. The results clearly show that using a purely dipolar potential, corresponding to an ideal dipole at the center of each ellipsoidal shell (Fig.~\ref{figure1}\textbf{f}), fails to capture the non-monotonic behavior of $\theta_{\mathrm{min}}$ against $\rho$ observed for two uniformly polarized ellipsoids (Fig.~\ref{figure1}\textbf{h}). The two-charge model (Fig.~\ref{figure1}\textbf{i}), however, captures this behavior, being similar to the corresponding curve for two uniformly polarized ellipsoids. (Note, however, that these results do not include the change in the local field due to mutual polarization.) In Fig.~\ref{figure1}\textbf{g}, we also discuss the model formerly proposed by Singh {\em et~al.}~\cite{singh}, which considers one uniformly polarized ellipsoid interacting with a point dipole inside an ellipsoidal shell. This model qualitatively captures the non-monotonicity of $\theta_{\mathrm{min}}$, but strongly overestimates the contact angles at minimum electrostatic energy for all aspect ratios. We furthermore notice that, for all four models, there is a distinct value of $\rho$ below which $\theta_{\mathrm{min}} = 0$, corresponding to the ``head-to-tail'' configuration preferred for dipolar spheres~\cite{dassanayake,hynninen}. The value of $\rho$ where $\theta_{\mathrm{min}}$ goes from zero to positive thus roughly marks the point where the effects of particle anisotropy will start dominating the observed structures; for the two-charge and uniformly polarized models, this transition occurs at aspect ratios of $\rho \approx 1.8$ and $\rho \approx 2.1$, respectively.

\section{Simulation details}\label{sec:simulations}

We performed Monte Carlo (MC) simulations of systems of monodisperse, hard ellipsoids in the canonical (constant $N,V,T$) ensemble using the MOLSIM package~\cite{Molsim}. Periodic boundary conditions were applied in all three dimensions, and the three box dimensions were held fixed at $L_x = L_y = 36 R_0$ in the $x$ and $y$ directions (perpendicular to the applied field) and $L_z = 60 R_0$ in the direction parallel to the field, where $R_0 \equiv [(3/4\pi) V_p]^{1/3}$ is the particle radius for $\rho = 1$, \emph{i.e.}, before the isochoric transformation from a sphere to an ellipsoid (see further the Model Description section). The particle long axis $a$ was fixed parallel to the $z$ axis (representing the direction of the external field), while the particle positions were evolved through single-particle trial translational moves. All simulations were run for $10^{7}$ MC cycles, where each cycle consists of one trial move per particle, and hard ellipsoid overlap was checked following Perram and Wertheim~\cite{perram}. The volume fractions of the simulations were fixed to $\phi=0.054$ unless otherwise stated. In addition to their excluded volume potential, particles interact through the electrostatic energy $U_{\mathrm{el}}$ obtained by a pairwise summation over the (unscreened) Coulombic interaction between all sites $i$ and $j$:

\begin{equation}
U_{\mathrm{el}} = \frac{1}{4 \pi \epsilon_{0} } \sum_{i} \sum_{j>i} \frac{q_{i} q_{j}}{r_{ij}}, \label{elec_p}
\end{equation}

\noindent where the sum runs over all charges $q_i$ in the system, excluding the interaction between sites on the same particle, and $r_{ij}$ is the separation between sites $i$ and $j$. The long-range part due to the periodic boundaries was included into $U_{\mathrm{el}}$ using the Ewald summation technique~\cite{FrenkelSmit}. 

To elucidate why, for large enough aspect ratios, we start observing 2D structures (sheets and tubes) instead of 3D crystals, we furthermore performed energy minimizations (\emph{i.e.}, corresponding to the limit $T \rightarrow 0$) as follows. The particles were arranged into the respective candidate structures (see Fig. \ref{figure2}) and their lattice parameters were expressed using two degrees of freedom chosen based on the symmetries of these crystal structures. The electrostatic energy of the lattice was then minimized by shrinking and expanding the structure through these two degrees of freedom, using Powell's method\cite{powell} by defining the maximum number of iterations and the desired fractional tolerance. As in the case of MC simulations, we used periodic boundary conditions and the long-range electrostatic interactions were handled using the Ewald summation technique.

\section{Simulation results}

\begin{figure*}[ht!]
\includegraphics[width=.7\textwidth]{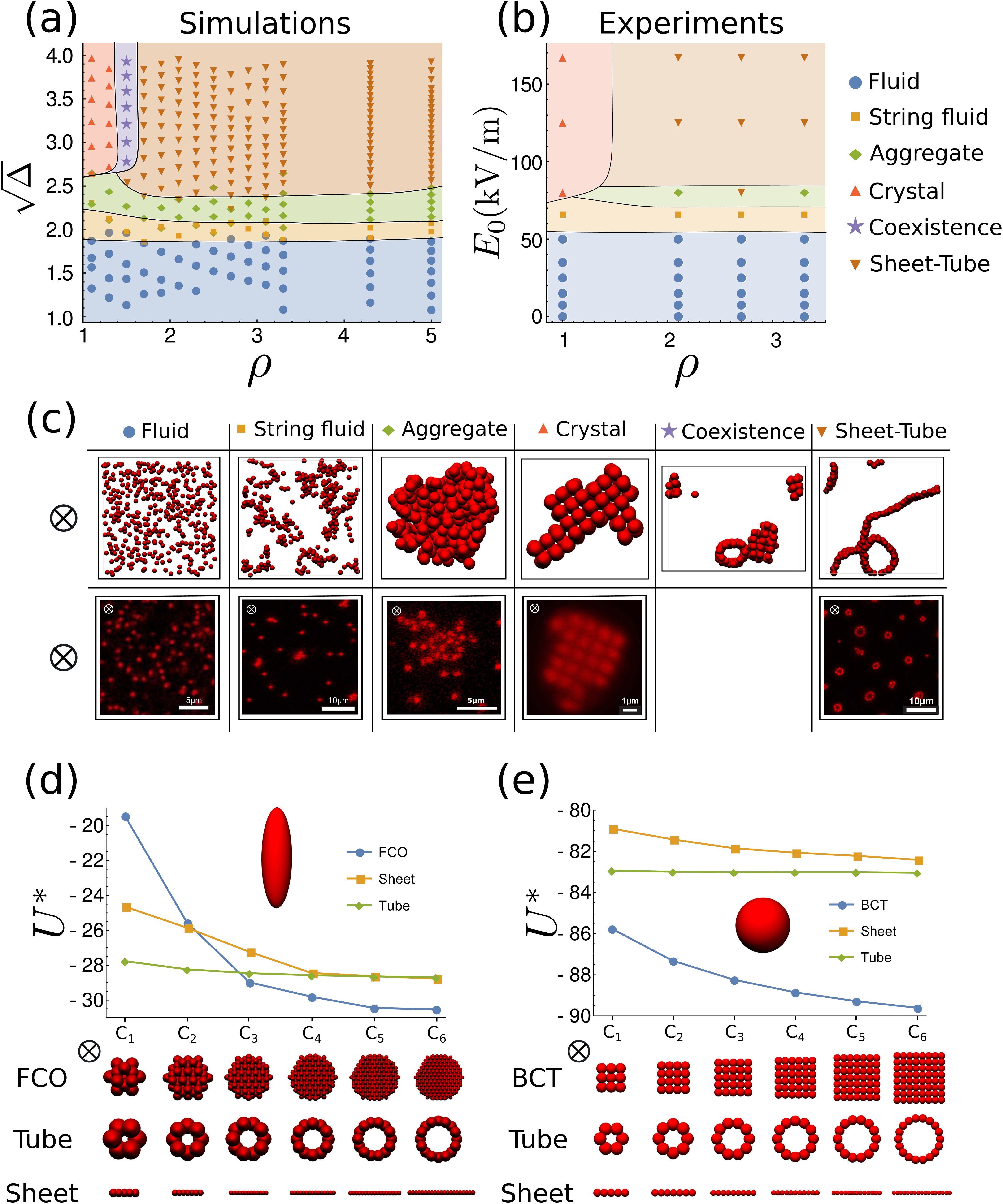}
\caption{State diagrams as a function of electric field strength and aspect ratio as obtained from (a) MC simulations at constant volume fraction $\phi = 0.054$ and (b) experiments on ellipsoidal colloids in an AC electric field at $\phi \approx 0.04$ (reproduced from Ref.~\cite{Jerome}). Panel (c) shows representative snapshots of the various states from simulations (top row) and experiments (bottom row), shown with the field direction perpendicular to the page. (d, e) Reduced electrostatic energy per particle ($U^* = U_{\mathrm{el}} / (Nk_{\mathrm{B}}T)$) for energy-minimized structures (BCT/FCO crystals, sheets, and tubes) of various sizes, as indicated, for (d) $\rho=3.3$ and (e) $\rho=1.01$. Note that the energy-minimized structures in (d, e) neglect the effect of entropy and thus correspond to the limit $T \rightarrow 0$, while the values of $qd$ and $T$ formally used to express $U^*$ are the same as for MC simulations at $\sqrt{\Delta} = 3.05$.} \label{figure2}
\end{figure*}

Having established the accuracy of our two-charge model, in order to study the self-assembly behaviour of the system we performed Monte Carlo simulations as described in Section \ref{sec:simulations}. We will present the simulation results in terms of the dimensionless electrostatic coupling parameter $\Delta$, defined by

\begin{equation}
\Delta = \frac{(qd)^2}{6 \epsilon_0 V_p k_{\mathrm{B}}T}, \label{energy_touch}
\end{equation}

\noindent where $k_{\mathrm{B}}$ is Boltzmann's constant and $T$ the absolute temperature. $\Delta$ quantifies the strength of the electrostatic coupling, and can be identified as the leading-order dipole-dipole coupling between two ellipsoids in a side-by-side configuration, divided by $\rho$ and normalized by $k_{\mathrm{B}}T$. As the physically relevant quantity is the ratio of $U_{\mathrm{el}}$ and $k_{\mathrm{B}}T$, rather than each quantity separately, the dimensionless quantity $\Delta$ is sufficient to describe the strength of electrostatic interactions and thus the physics of the system. In experiments, ellipsoids were found to align at essentially the same field strength irrespective of $\rho$. In the following, we will thus assume that the dipole moment $qd$ is independent of $\rho$ at fixed external field, implying that $\Delta$ is also constant, since the particle volume $V_p$ was not changed when changing the particle aspect ratio. We furthermore have that $\Delta$ is directly proportional to $E_0^2$ as long the induced dipole moment scales linearly with $E_0$; the MC simulations thus do not explicitly include any coupling to an external field. 

Figure \ref{figure2} shows state diagrams as obtained from MC simulations and compared with previously reported experiments on ellipsoidal particles at low volume fractions~\cite{Jerome}. In spite of the approximations of the model, the similarity between the observed structures is striking, indicating that the two-charge model indeed captures the important interactions present in the experiments. At very low electrostatic coupling (low $\sqrt{\Delta}$ or $E_0$), the system consists of a fluid of free particles aligned with the field direction. At slightly higher coupling strengths ($1.9 \lesssim \Delta \lesssim 2.4$), these start to aggregate into a fluid of short, 1-dimensional strings, followed by a region where 3-dimensional amorphous aggregates form. At even higher coupling strengths ($\Delta \gtrsim 2.4$), and $\rho \geq 1.5$, we then observe stable phases of coexisting 2-dimensional sheets and microtubules. At lower aspect ratios, the stable state is instead a body-centered tetragonal (BCT or BCO) crystal, as has been observed and theoretically predicted several times before for spherical particles in an electric field~\cite{dassanayake,yethiraj,hynninen,Sofi,Mohanty2015PRX}. The transition from crystals to sheets and tubes approximately coincides with the aspect ratio ($\rho \approx 1.5$) above which the head-to-tail configuration of two particles is no longer favorable, with a small region where we observe coexistence between crystallites and sheets or tubes. Interestingly, such coexistence was previously observed for slightly anisotropic bowl-shaped particles with an effective aspect ratio of about 1.4\cite{Jerome2}. 

We can obtain an approximate mapping between the coupling strength $\Delta$ and the field strength $E_0$ by assuming that the particles are polarized solely along their long axis, and considering that particles then are expected to align at an interaction energy of $\frac{1}{2} \mu_{\mathrm{ind}} E_0 \approx k_B T$, where $\mu_{\mathrm{ind}}$ is the induced dipole moment.\cite{okonski} In experiments, we observe alignment at $E_0 \approx 25$ kV/m at $T = 20^\circ$C, yielding $\mu_{\mathrm{ind}} = qd \approx 3.2 \times 10^{-25}$ Cm at this field strength. We then use the measured hydrodynamic radius of the spherical particles ($R_H = 537$ nm\cite{Jerome}) to define $V_p$, which through Eq. \eqref{energy_touch} yields $\sqrt{\Delta} \approx 0.9$. Particle assembly is furthermore observed in experiments for $E_0 \gtrsim 50$ kV/m, which thus corresponds to $\sqrt{\Delta} \gtrsim 1.7$, in excellent agreement with the observed onset of string formation in simulations (see Fig. \ref{figure2}a).

We note that, in our simulations, sheets and tubes often coexist, and several different realizations of the same simulation conditions might give either (or both) structures. To shed light on the transition from $3{\rm D}$ (crystals) to $2{\rm D}$ (sheets and tubes) structures by increasing the aspect ratio $\rho$, in Fig.~\ref{figure2}\textbf{d,e} we analyze the electrostatic energy, \emph{i.e.}, corresponding to the limit $T \rightarrow 0$, of these different structures for two different values of $\rho$. Figure \ref{figure2}\textbf{d} shows that the face-centered orthorhombic (FCO) crystal structure is higher in energy than the sheet and tube structures for small aggregate sizes, which supports the absence of crystal formation for high-aspect ratio ellipsoids at low $\phi$. In contrast, in the near-spherical case ($\rho = 1.01$, Fig.~\ref{figure2}\textbf{e}) the BCT crystal structure is the energetically favored state for all structure sizes. This observation, together with the observation that the FCO crystalline state appears to be the state of minimum energy for large aggregates (corresponding to the thermodynamically favored phase for strong electrostatic coupling), indicates that the sheet and tube structures are likely to be the result of kinetic trapping in a local free energy minimum. Furthermore, the fact that tubes are less frequently observed in simulations than in experiments (see snapshots in Fig.~\ref{figure2}\textbf{c}), while large sheets are not observed in experiments, is likely due to the fact that the MC chain consists solely of single-particle translational moves, which will not accurately sample the collective displacements needed for tubular formation.

\begin{figure*}[t]
\includegraphics[width=.45\textwidth]{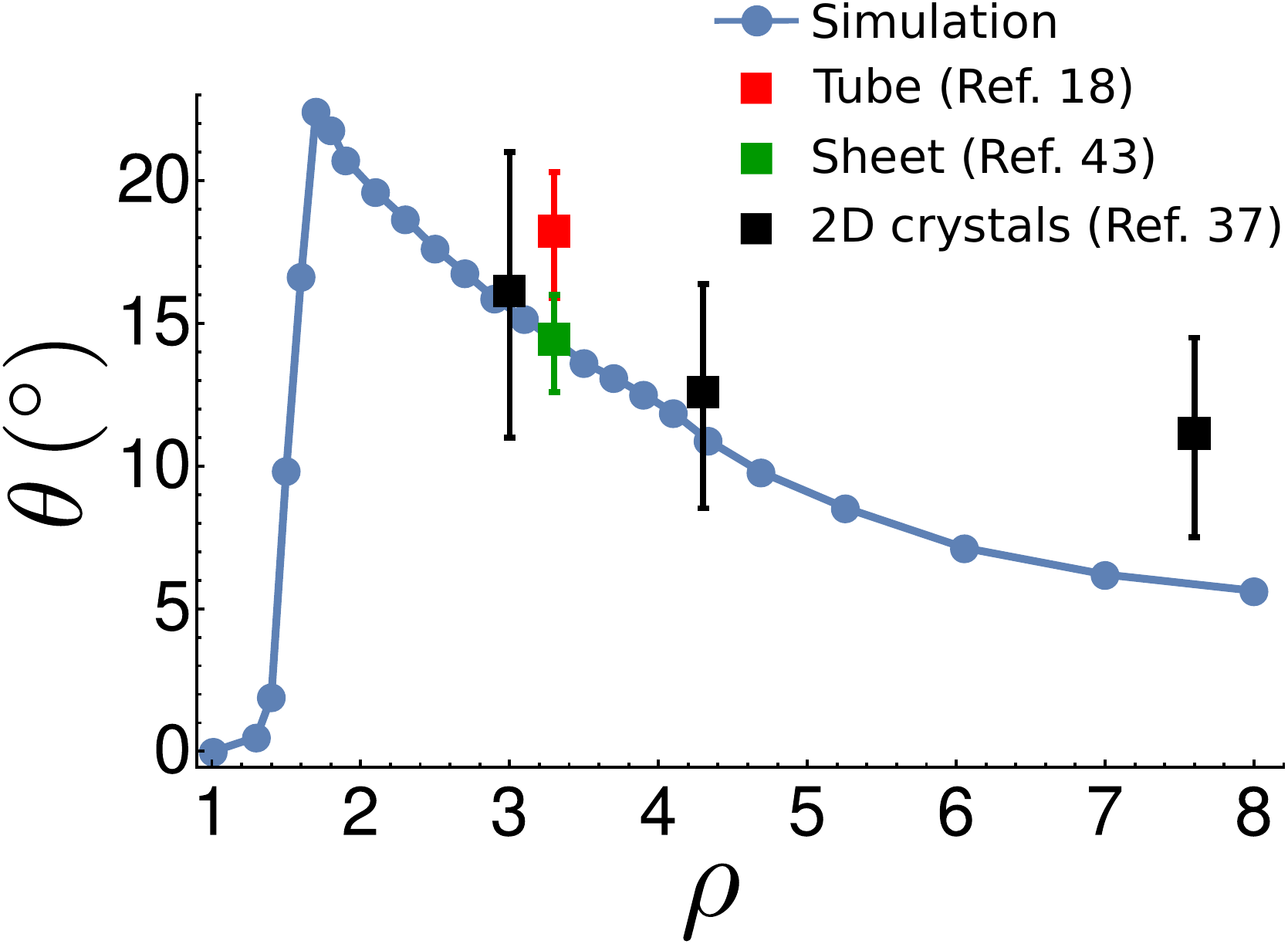}
\caption{Ensemble average of the nearest-neighbor angle $\theta$ as a function of $\rho$ for $\Delta=3.05$ and $\phi=0.054$ compared to the experimentally obtained values reported by Singh \emph{et al.}~\cite{singh} and for ellipsoidal microgel particles with $\rho=3.3$ assembled in tubes~\cite{Jerome} or in sheets~\cite{JeromeN}.} \label{figure3}
\end{figure*}

\begin{figure*}[ht!]
\includegraphics[width=.55\textwidth]{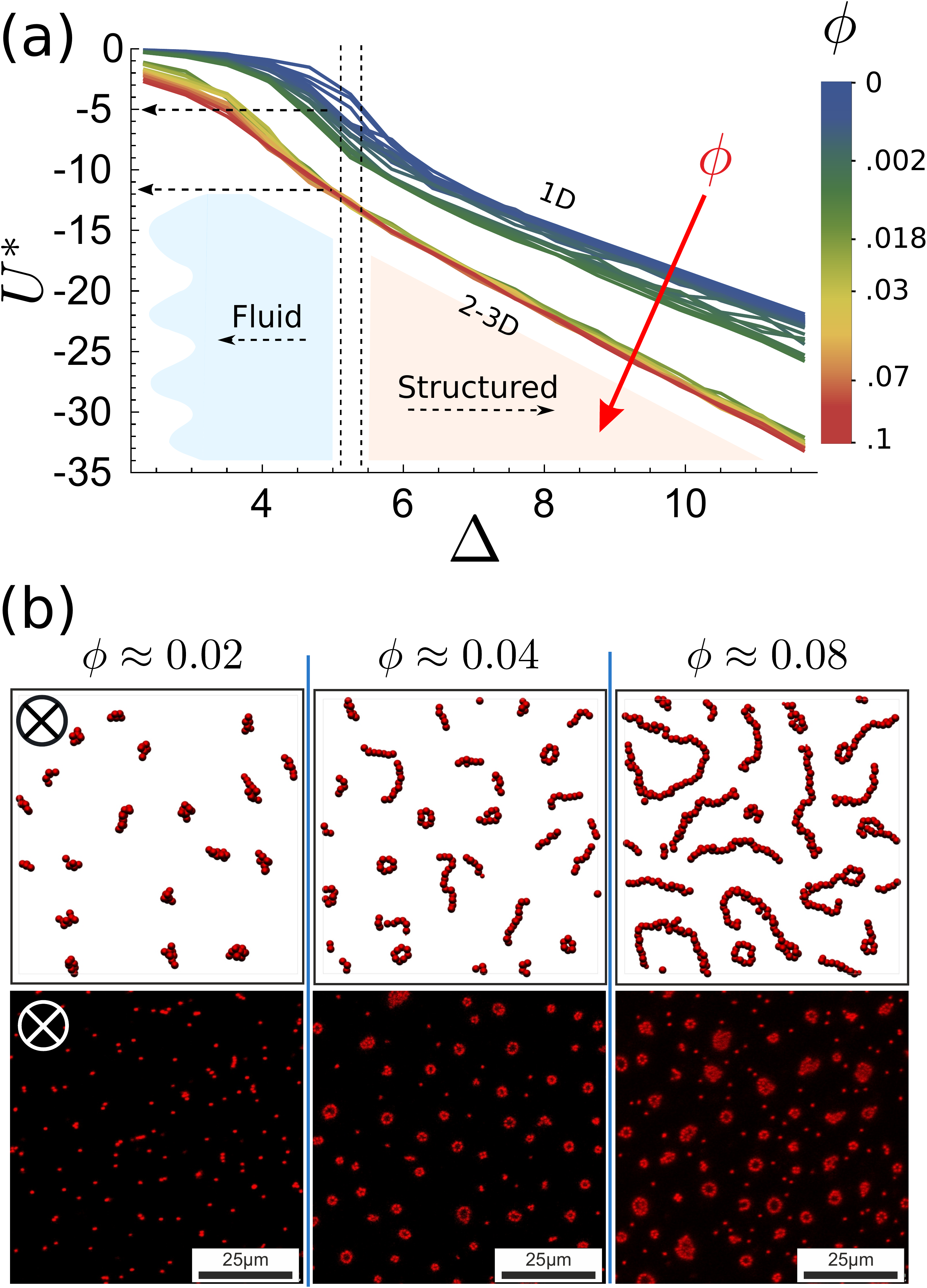}
\caption{(a) Ensemble averaged electrostatic energy per particle $U^* \equiv \langle U_{\mathrm{el}} \rangle / (Nk_{\mathrm{B}}T)$ as a function of the coupling parameter $\Delta$ for several values of the particle volume fraction $\phi$, as indicated, at constant aspect ratio $\rho=3.3$; note the transition at $\phi \approx 0.018$ from 1${\rm D}$ to 2-3${\rm D}$ structures. (b) Corresponding snapshots from simulations with $\sqrt{\Delta} = 2.85$ $(\Delta = 8.12)$ (top) and experiments with $E_0 = 167$ kVm$^{-1}$ (bottom) at various values of $\phi$, as indicated.} \label{figure4}
\end{figure*}

Next, the nearest-neighbor angle $\theta$ as function of $\rho$ obtained at $\Delta=3.0$ and $\phi=0.054$ is shown in Fig.~\ref{figure3}, and is compared to the experimental values reported by Singh $\emph{et al}$.\cite{singh} for polystyrene ellipsoids and for composite microgels with $\rho=3.3$ assembled in tubes\cite{Jerome} or sheets\cite{JeromeN} (Note that $\theta$ could not be determined for the other aspect ratios due to the larger mobility of the tubular assemblies in the image plane). The comparison demonstrates a good agreement between experiments and two-charge model simulations for $\rho<5$. Furthermore, the $\theta$-values for intermediate aspect ratios are considerably higher than the corresponding energy-minimized two-body values $\theta_\mathrm{min}$ in Fig. \ref{figure1}\textbf{e}, due to the effects of many-body interactions and non-zero temperature in the simulations.

Finally, in Fig.~\ref{figure4}\textbf{a}, we assess the effect of varying the particle volume fraction $\phi$ of the system, by analyzing the average electrostatic energy per particle $U^* \equiv \langle U_{\mathrm{el}} \rangle / (Nk_{\mathrm{B}}T)$ as a function of $\Delta$ for a range of $\phi$ values. A clear transition from 1${\rm D}$ (strings) to 2${\rm D}$ and 3${\rm D}$ (aggregates, sheets and tubes) structures is seen around $\phi \approx 0.018$, visible as a sharp increase in the slope of $U^*$ versus $\Delta$ due to the increased number of nearest neighbours in the higher-dimensional structures. Finally, by following a single curve (\emph{i.e.}, for a constant value of $\phi$) one can identify two coupling regimes: a low-coupling one for $\Delta \leq 5$ dominated by fluid-like structures, and a high-coupling regime for $\Delta \geq 6$, where solid-like phases are formed, with a narrow crossover regime where amorphous aggregates dominate the structures. The electrostatic interaction energies required to form solid-like aggregates in 1${\rm D}$ is $U^* \approx -5$ and approximately twice in 2/3${\rm D}$ ($U^* \approx -12$), due to an increasing number of neighboring particles. Snapshots from simulations and experiments are shown in Fig.~\ref{figure4}\textbf{b} to illustrate the $\phi$-dependence of the observed structures.\\

\section{Conclusions}

In this study, we have presented Monte Carlo simulations of a two-charge model of polarizable ellipsoidal colloids in an external electric field. The simulated state diagram at low volume fractions is qualitatively very similar to the one observed in experiments, exhibiting a rich phase behavior comprising strings, sheets, tubes, and crystals. The fact that the experimental structures are accurately captured by our simplified model further indicates that the effect of including many-body interactions due to varying local electric fields will at most have a quantitative effect on the state behaviour (see further~\cite{dassanayake,rotunno}), although it might partially explain the fact that tubes occur more frequently in experiments than in simulations. This is particularly encouraging given the high computational cost of such many-body interactions, which would require an iterative procedure to obtain the charge distribution on every MC step. 

Our results furthermore show that the particle anisotropy as measured through the aspect ratio $\rho$ is a key parameter in determining the transition from crystals ($\rho \leq 1.5$) to sheets and tubes ($\rho \geq 1.5$), due to the minimum energy configuration of adjacent particles shifting from the head-to-tail configuration favored for small aspect ratios to association of two adjacent particles at a non-zero angle for $\rho \geq 1.5$. This behavior is distinctly different from that observed for polarized spherocylinders, where the head-to-tail configuration continues to be favored even for large aspect ratios at the two-particle level, thus leading to a different state diagram~\cite{rotunno,troppenz,liu}. Our results therefore highlight how subtle details of anisotropic steric interactions can be used in conjunction with long-range anisotropic interactions to yield new routes to directed self-assembly at the mesoscopic scale.

\begin{acknowledgement} 
The authors acknowledge Jan Vermant and Patrick Pfleiderer for the processing of the ellipsoidal particles. We would like to thank Jan~K.G. Dhont and H\aa{}kan Wennerstr\"{o}m, for helpful discussions and A.~A. would like to thank Anita Goebel and Parisa Azari for their help. This work was financed through the Knut and Alice Wallenberg Foundation (project grant KAW 2014.0052) and the European Research Council (ERC-339678-COMPASS). J.~S. ackowledges funding from the Swedish Research Council (2015-05449).
\end{acknowledgement}

\bibliography{ref}

\end{document}